\documentclass[%
preprint,
superscriptaddress,
 amsmath,amssymb,
 aps,
]{revtex4-2}

\usepackage[english]{babel} 
\usepackage[T1]{fontenc}
\usepackage[ansinew]{inputenc}

\usepackage{amsmath}
\usepackage{amsthm}
\usepackage{amsfonts}
\usepackage{bbold}
\usepackage{xcolor}
\usepackage{soul}

\usepackage{graphicx}
\usepackage{dcolumn}
\usepackage{bm}
\newcommand{\B}{\mathbf{B}}
\newcommand{\M}{\mathbf{M}}
\renewcommand{\H}{\mathbf{H}}
\newcommand{\Ha}{\mathbf{H}_{\rm a}}
\newcommand{\NN}{\mathbb{N}}

\renewcommand{\SS}{\mathbb{S}}
\newcommand{\QQ}{\mathbb{Q}}

\begin{document}

\preprint{v040 - \today}

\title{Modelling magnetically-levitated superconducting ellipsoids, cylinders and cuboids for quantum magnetomechanics}
\author{Natanael Bort-Soldevila}
\thanks{These two authors contributed equally to this work.}
\author{Jaume Cunill-Subiranas}
\thanks{These two authors contributed equally to this work.}
\author{Nuria Del-Valle}
\affiliation{Departament de F\'isica, Universitat Aut\`onoma de Barcelona, 08193 Bellaterra, Barcelona, Spain}
\author{Witlef Wieczorek}
\affiliation{Department of Microtechnology and Nanoscience (MC2), Chalmers University of Technology, SE-412 96 Gothenburg, Sweden}
\author{Gerard Higgins}
\affiliation{Department of Microtechnology and Nanoscience (MC2), Chalmers University of Technology, SE-412 96 Gothenburg, Sweden}
\affiliation{Institute for Quantum Optics and Quantum Information (IQOQI), Austrian Academy of Sciences, A-1090 Vienna, Austria}
\author{Michael Trupke}
\affiliation{Institute for Quantum Optics and Quantum Information (IQOQI), Austrian Academy of Sciences, A-1090 Vienna, Austria}
\affiliation{Vienna Center for Quantum Science and Technology, Department of Physics, University of Vienna, A-1090 Vienna, Austria}
\author{Carles Navau}
\email{carles.navau@uab.cat}
\affiliation{Departament de F\'isica, Universitat Aut\`onoma de Barcelona, 08193 Bellaterra, Barcelona, Spain}


\begin{abstract}
We theoretically investigate the properties of magnetically-levitated superconducting rotors confined in anti-Helmholtz traps, for application in magnetomechanical experiments.
We study both the translational modes and a librational mode.
The librational mode gives an additional degree of freedom that levitated spheres do not have access to.
We compare rotors of different shapes: ellipsoids, cylinders and cuboids.
We find that the stable orientations of the rotors depend on the rotors' aspect ratios.
\end{abstract}


\maketitle

\section{Introduction \label{sec.intro}}
When an external uniform magnetic field $\H_{\rm a}$ is applied over a magnetic material with the shape of an ellipsoid, it magnetizes with a uniform magnetization $\M$ \cite{Chen1991}. If the magnetic material is linear, $\B=\mu \H$ inside the ellipsoid ($\mu$ is the constant permeability), and the magnetization is related to the internal field through the susceptibility constant $\chi$ as $\M=\chi \H$. Thus, $\chi=(\mu/\mu_0)-1$ being $\mu_0$ the vacuum permeability. 

When $\mu\rightarrow 0$ ($\chi\rightarrow-1$) the material behaves as a perfect diamagnet ($\B=0$ inside). 
A superconductor can be modeled as a perfect diamagnet, although there are extra properties, i.e. flux quantization, that are not captured by just $\chi\rightarrow-1$. On the other limit, when $\mu\rightarrow\infty$ ($\chi\rightarrow\infty$) we would be considering a perfectly soft ferromagnet ($\H=0$ inside).

The demagnetizing factors are commonly used in magnetic experiments. Indeed, the experimentalist controls the applied field, while the intrinsic properties one usually wants to measure (i.e., the susceptibility) depend on the internal field. Both are related through the demagnetizing factors.

We focus our study on levitation systems in the context of quantum magnetomechanics. Indeed, the levitation of small diamagnetic microparticles in specially designed magnetic traps has been proposed as the experimental platform to perform ground-state cooling of the center-of-mass degrees of freedom of the superconductor \cite{Romero-Isart2012,Cirio2012,Gutierrez2023}. Different magnetic traps have been proposed, such as surface superconductors \cite{Timberlake2019,Prat-Camps2017,Vinante2020}, quadrupolar magnets \cite{Slezak2018}, chip-based multi-winding planar coils \cite{Gutierrez2023}, 3D arranged coils \cite{Cirio2012}, or anti-Helmholtz coils (AHC) \cite{Romero-Isart2012}, to cite some. Stable levitation of lead microspheres ($\simeq 50$ $\mu$m in diameter, 700 ng of mass) in anti-Helmholtz traps has been recently demonstrated \cite{Gutierrez2020,Gutierrez2023}. Using a similar system (with elliptical coils to break the $xy$-axis symmetry) the center-of-mass motion of levitated lead-tin spheres (now of $\simeq 100$ $\mu$m in diameter) have been feedback controlled after coupling with a superconducting quantum interference device achieving quality factors up to 2.6$\times 10^7$ \cite{Hofer2023}.

As for the applied field, we shall consider AHC generating the applied fields. In these levitation experiments, the applied field is not uniform and the use of demagnetizing factors should be carefully considered. At the equilibrium position, the applied field is quadrupolar with a value of zero at the center. The assumption of a uniform applied field does not apply. However, we will see how to take profit of the demagnetizing factors to account for the restoring force and obtain useful expressions for the trapping frequency.

As for the levitated particle, not only the size, but its shape can play an important role, since the demagnetizing fields can change drastically the total field it experiences. Although the spherical shape is the most used in magnetomechanics levitation experiments \cite{Gutierrez2022}, finding a perfect sphere is complicated and the studies of how the shape affects the levitation are important. The nonsphericity lifts some of the symmetries and some vibrational modes become non-degenerate. Moreover, non-spherical particles exhibit torques in the non-uniform magnetic field which lead to librational modes as additional degrees of freedom \cite{Zielimode2023}. Ellipsoids represent a good approximation for many other shapes ranging from long cylinders to disks (and including the sphere). Although the levitation of superconducting rings (including flux quantization) has been considered \cite{Navau2021,Gutierrez2022}, there is a lack of work devoted to studying the superconductor's shape influence.

In this work, we aim to conduct an investigation of the levitation characteristics of magnetic particles in a quadrupolar (anti-Helmholtz) magnetic trap. We mainly focus on superconducting ellipsoids, but some comparison with cylinders and cuboids is also presented. For the ellipsoids, we derive analytic expressions for the translational trapping frequencies as a function of the principal axis lengths. For the librational frequencies, as well as for the comparison with cylinders and cuboids, we use finite element simulations.

The paper is structured as follows. In section \ref{sec.ext.susc.ellip} we derive the expressions for the magnetization of a generic ellipsoidal linear magnetic material in the presence of a uniform applied field. We find the external susceptibility matrix that relates the magnetization of the ellipsoid with the external applied field. In section \ref{sec.forcesandtorques}, we consider a quadrupolar field and find how the forces and torques over the levitated ellipsoid near its center can be found. We evaluate the translational and librational torques for this trap. In sections \ref{sec.comparison} and \ref{subsec.torques}, numerical results comparing translational and librational frequencies, respectively, of the trap for levitated superconducting ellipsoids, cylinders, and cuboids are presented. We finish with conclusions in section \ref{sec.conclusions}.

\begin{figure}[t]
    \centering
    \includegraphics[width=0.48\textwidth]{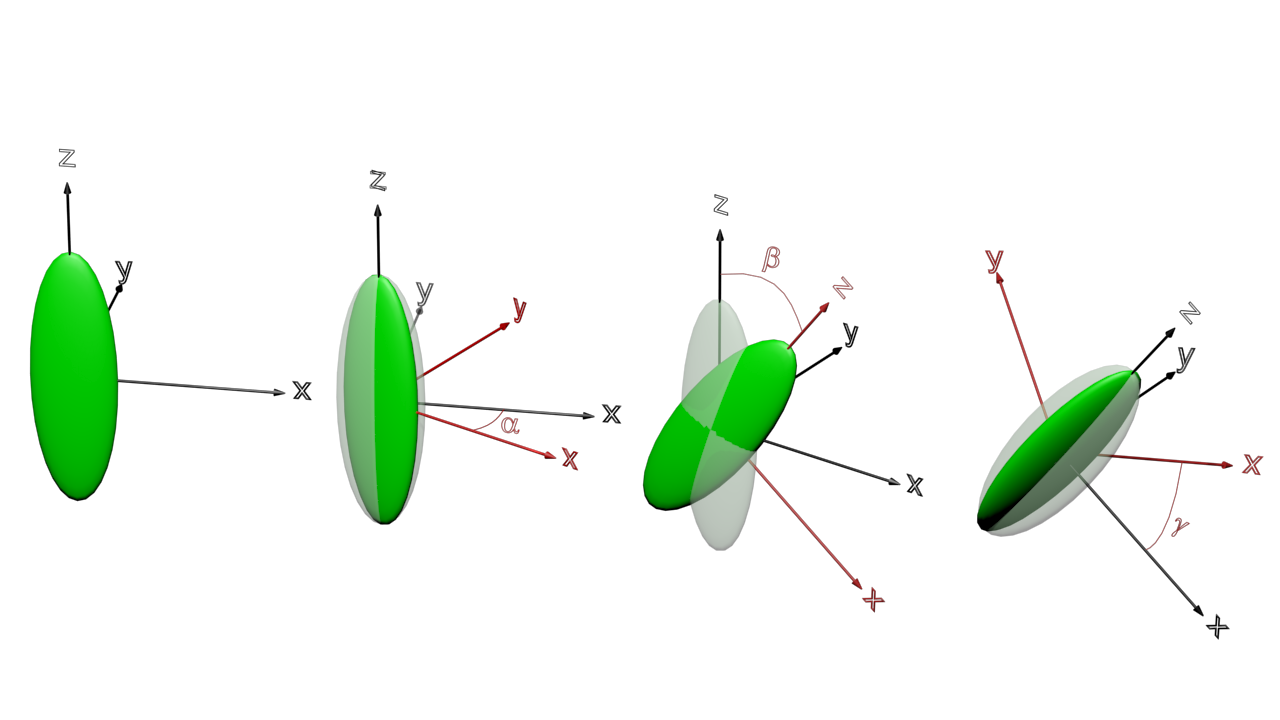}
    \caption{Sketch of a general rotation of an ellipsoid. The axes after rotating the ellipsoid are red, and the ellipsoid once rotated is intense green. When $a=b$, the $\alpha$ angle can be skipped and the $\gamma$ angle is irrelevant (since $x$ and $y$ are identical). Thus, when $a=b$, only $\beta$ is a relevant rotation angle.}
    \label{fig:rotation}
\end{figure}

\section{The external susceptibility for ellipsoids}
\label{sec.ext.susc.ellip}

We consider a linear magnetic material with susceptibility $\chi$ (an ideal superconductor would be $\chi=-1$) and with the shape of an ellipsoid with principal semiaxes $a$, $b$, and $c$.
The induction field $\B$ inside the ellipsoid is related to the magnetization $\M$, the magnetic field $\H$, the applied  field $\H_{\rm a}$ (uniform, in this section), and the demagnetizing field $\H_{\rm d}$, through
\begin{equation}
\B=\mu_0 (\M + \H)=\mu_0 (\M + \H_{\rm a} + \H_{\rm d}).
\label{eq:def1}
\end{equation}

The magnetization of the ellipsoid is related to the demagnetizing field through the diagonal demagnetizing factor matrix $\NN$ when both $\M$ and $\H_{\rm d}$ components are expressed as projections along the principal axes ($\hat{a}$, $\hat{b}$, and $\hat{c}$) of the ellipsoid. In this case, one has $\H_{\rm d}'=-\NN \M'$ where the primes ($'$) indicate projections along principal axes and
\begin{equation}
\NN=
\begin{pmatrix}
    N_a & 0 & 0 \\
    0 & N_b & 0 \\
    0 & 0 & N_c
\end{pmatrix}.
\label{eq:NN}
\end{equation}
The demagnetizing factors $N_a$, $N_b$, and $N_c$ satisfy $N_a+N_b+N_c=1$. Using Eq.~\eqref{eq:def1}, 
\begin{equation}
\M' = \chi (\mathbb{1}+ \chi \NN)^{-1} \H_{\rm a}'\, .
\label{eq:external}
\end{equation}

We describe the orientation of the ellipsoid using the Euler angles $zyz$ representation of rotation defined as follows (see Fig. \ref{fig:rotation}): (i) Initially the ellipsoid's principal axis coincides with the Cartesian axis, $\hat{a}=\hat{x}$, $\hat{b}=\hat{y}$, and $\hat{c}=\hat{z}$; (ii) first, rotate the ellipsoid an angle $\alpha$ with respect to the $z$ axis; (iii) second, rotate the (already rotated) ellipsoid an angle $\beta$ with respect to the rotated $y$ axis; (iv) finally, rotate the (already rotated) ellipsoid an angle $\gamma$ with respect to the rotated $z$ axis.

Any rotation within this representation can be described by ($\alpha,\beta,\gamma$). The rotation matrix $\QQ$ is 
\begin{widetext}
\begin{equation}
\QQ=
\begin{pmatrix}
 \cos \alpha \cos \beta \cos \gamma-\sin \alpha \sin \gamma & -\cos \gamma
   \sin \alpha-\cos \alpha \cos \beta \sin \gamma & \cos \alpha \sin \beta
   \\
 \cos \beta \cos \gamma \sin \alpha+\cos \alpha \sin \gamma & \cos \alpha
   \cos \gamma-\cos \beta \sin \alpha \sin \gamma & \sin \alpha \sin \beta
   \\
 -\cos \gamma \sin \beta & \sin \beta \sin \gamma & \cos \beta \\
\end{pmatrix}
\label{eq:QQ}
\end{equation}
\end{widetext}
We use this $\QQ$ matrix to change the coordinates of the vectors. Actually, we have $\H_{\rm a}' =\QQ \H_{\rm a}$ or $\H_{\rm a}=\QQ^T \H_{\rm a}'$, where $^T$ indicates the transposed matrix. From Eq.~\eqref{eq:external} we can write the magnetization vector, expressed in Cartesian coordinates as a function of the \textit{applied} field, also expressed in Cartesian coordinates:
\begin{equation}
\M = \SS \H_{\rm a}\, ,
\label{eq:MVsHa}
\end{equation}
where we have defined the \textit{external susceptibility matrix} $\SS$ as
\begin{equation}
\SS = \chi \QQ^T (\mathbb{1}+\chi\NN)^{-1} \QQ\, .
\label{eq:SSdef}
\end{equation}

The importance of Eq.~\eqref{eq:MVsHa} relies on that it gives the magnetization of an arbitrarily rotated linear, homogeneous, isotropic, and magnetic ellipsoid as a consequence of a uniform applied field, directly as a function of the applied field (not the internal field $\H$). The obtained matrix $\SS$ is general and can be applied to any ellipsoid as long as the applied field is considered uniform through the ellipsoid. The explicit general expression is cumbersome although straightforward from Eqs. \eqref{eq:NN}, \eqref{eq:QQ}, and \eqref{eq:SSdef}. 

\subsection{Spheroid, $a=b$}
As a particular case of the general treatment done above, we consider here that the ellipsoid has two identical principal axes, $a=b$. Then, $N_a=N_b\equiv N_{ab}$. Considering the applied field along the $z$ direction, any rotation is described only by $\beta$. 
That is, we are rotating the spheroid about the $y$-axis at an angle $\beta$. In this case, the external susceptibility matrix reduces to
\begin{widetext}
\begin{equation}
\SS=
\begin{pmatrix}
 \frac{\chi  ((N_{ab}+N_{c}) \chi -(N_{ab}-N_{c}) \cos (2 \beta ) \chi +2)}{2 (N_{ab} \chi +1) (N_{c} \chi +1)} & 0 &
   \frac{(N_{c}-N_{ab}) \chi ^2 \cos \beta \sin \beta}{(N_{ab} \chi +1) (N_{c} \chi +1)} \\
 0 & \frac{\chi }{N_{ab} \chi +1} & 0 \\
 \frac{(N_{c}-N_{ab}) \chi ^2 \cos \beta \sin \beta}{(N_{ab} \chi +1) (N_{c} \chi +1)} & 0 & \frac{\chi 
   ((N_{ab}+N_{c}) \chi +(N_{ab}-N_{c}) \cos (2 \beta ) \chi +2)}{2 (N_{ab} \chi +1) (N_{c} \chi +1)}
\end{pmatrix}.
\label{eq:SSspheroid}
\end{equation}  
\end{widetext}

\begin{figure}[t]
    \centering
        \includegraphics[width=0.48\textwidth]{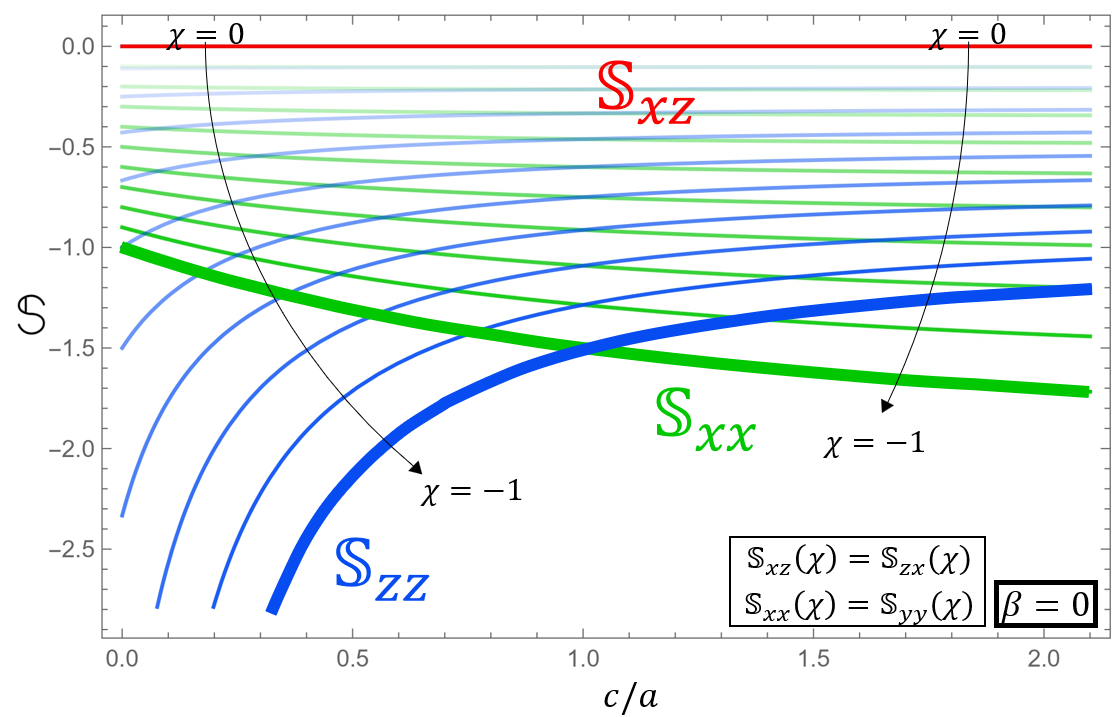}
        \caption{$\SS_{xx}$ (green), $\SS_{zz}$ (blue), and $\SS_{xz}$ (red) matrix elements of a spheroid, as a function of $c/a$ for different values of $\chi$, from 0 to $-1$ in intervals of $-0.1$ (following the arrows). The thickest lines correspond to curves with $\chi=-1$. In this figure, $\beta=0$. Note that the rest of the matrix is given by $\SS_{zx}(\chi)=\SS_{xz}(\chi)$, $\SS_{xx}(\chi)=\SS_{yy}(\chi)$, and $\SS_{xy}=\SS_{yx}=\SS_{yz}=\SS_{zy}=0$.}
        \label{fig:generalS-chi}
\end{figure}

Importantly, for the spheroid, the demagnetizing factors of $\NN$ have been analytically found \cite{Stoner1945,Takahashi2017} as 
\begin{widetext}
\begin{eqnarray}
    N_c&=& 
    \begin{cases}
    \frac{1}{1-\left(\frac{c}{a}\right)^2}\left(1-\frac{\frac{c}{a}}{\sqrt{1-\left(\frac{c}{a}\right)^2}} \arccos{\frac{c}{a}}\right), \textrm{ if }c<a, \\
    \frac{1}{\left(\frac{c}{a}\right)^2-1}\left(\frac{\frac{c}{a}}{\sqrt{1-\left(\frac{c}{a}\right)^2}}\ln\left(\frac{c}{a}+\sqrt{\left(\frac{c}{a}\right)^2-1} \right) -1 \right), \textrm{ if }c>a, 
    \end{cases} 
    \\
    N_{ab}&=&\frac{1}{2}(1-N_c)\, .
\end{eqnarray}
\end{widetext}
In Figs. \ref{fig:generalS} and \ref{fig:generalS-chi} we plot the values of different components of the $\SS$ matrix as a function of $c/a$ for several values of $\beta$ and $\chi$. When $\beta=0$ and when $\beta=\pi/2$ one obtains, respectively,
\begin{eqnarray}
\SS(\beta=0)=
\begin{pmatrix}
 \frac{\chi }{N_{ab} \chi +1} & 0 & 0 \\
 0 & \frac{\chi }{N_{ab} \chi +1} & 0 \\
 0 & 0 & \frac{\chi }{N_c \chi +1} 
\end{pmatrix},
\label{eq:DDnorot0}\\
\SS(\beta=\pi/2)=\begin{pmatrix}
 \frac{\chi }{N_c \chi +1}  & 0 & 0 \\
 0 & \frac{\chi }{N_{ab} \chi +1} & 0 \\
 0 & 0 & \frac{\chi }{N_{ab} \chi +1}
\end{pmatrix}.
\label{eq:DDnorotpi2}
\end{eqnarray}
For spheres, $N_c=N_{ab}=1/3$, the rotation matrix is the identity matrix, and the external susceptibility tensor becomes
\begin{equation}
    \SS = \frac{3\chi}{3+\chi}\mathbb{1}.
    \label{eq.SSsphere}
\end{equation}
For the ideal superconducting sphere ($\chi=-1$), we have $\SS = -(3/2)\mathbb{1}$. 

\section{Forces and torques over the ellipsoid in a quadrupolar external field}
\label{sec.forcesandtorques}

\subsection{Anti-Helmholtz coil  field}
\label{subsec.AHCoils}

One of the common traps for quantum magnetomechanics experiments is the AHC trap. It consists of two coaxial coils of radius $R$ separated by a distance $R$. A current $I$ circulates through the coils in the opposite direction. 
The origin of coordinates is located on the coaxial axis of the coils, at the equidistant point between all the points of both coils.
The field created by the AHC at positions ${\bf r}=(x,y,z)$ close to the origin ($|{\bf r}|\ll R$) is a quadrupolar field that can be written as
\begin{equation}
\Ha= \frac{H_0}{R} (-x {\bf\hat{x}}-y{\bf\hat{y}}+2z{\bf\hat{z}})\, ,
\label{eq:AHC}
\end{equation}
where $H_0\equiv\frac{24I}{25\sqrt{5}R}$. 

\begin{figure}[t]
    \centering
        \includegraphics[width=0.48\textwidth]{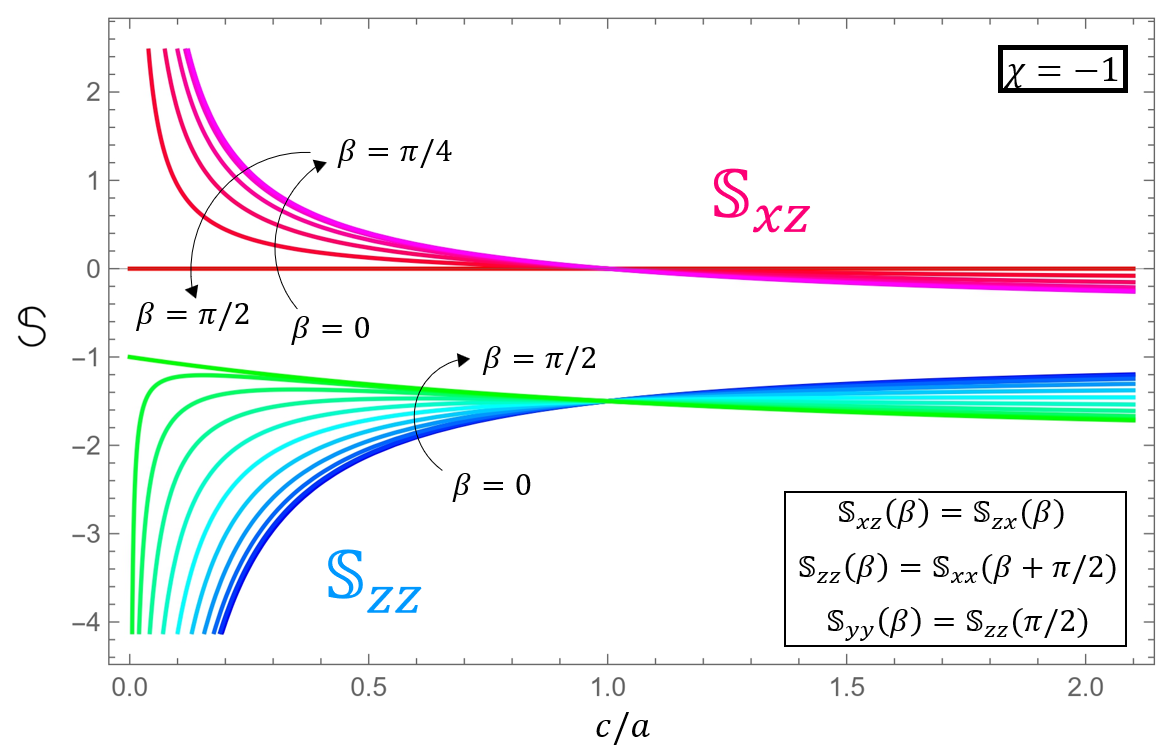}
        \caption{$\SS_{zz}$ (blue-green) and $\SS_{xz}$ (red-purple) matrix elements of a spheroid, as a function of $c/a$ for different values of $\beta$ from 0 to $\pi/2$ in intervals of $\pi/20$ (following the arrows). In this figure, $\chi=-1$. Note that the rest of the matrix is given by $\SS_{zx}(\beta)=\SS_{xz}(\beta)$, $\SS_{zz}(\beta)=\SS_{xx}(\beta+\pi/2)$, $\SS_{yy}(\beta)=\SS_{xx}(0)=\SS_{zz}(\pi/2)$, and $\SS_{xy}=\SS_{yx}=\SS_{yz}=\SS_{zy}=0$.}
        \label{fig:generalS}
\end{figure}

\begin{figure}[t]
    \centering        
    \includegraphics[width=0.48\textwidth]{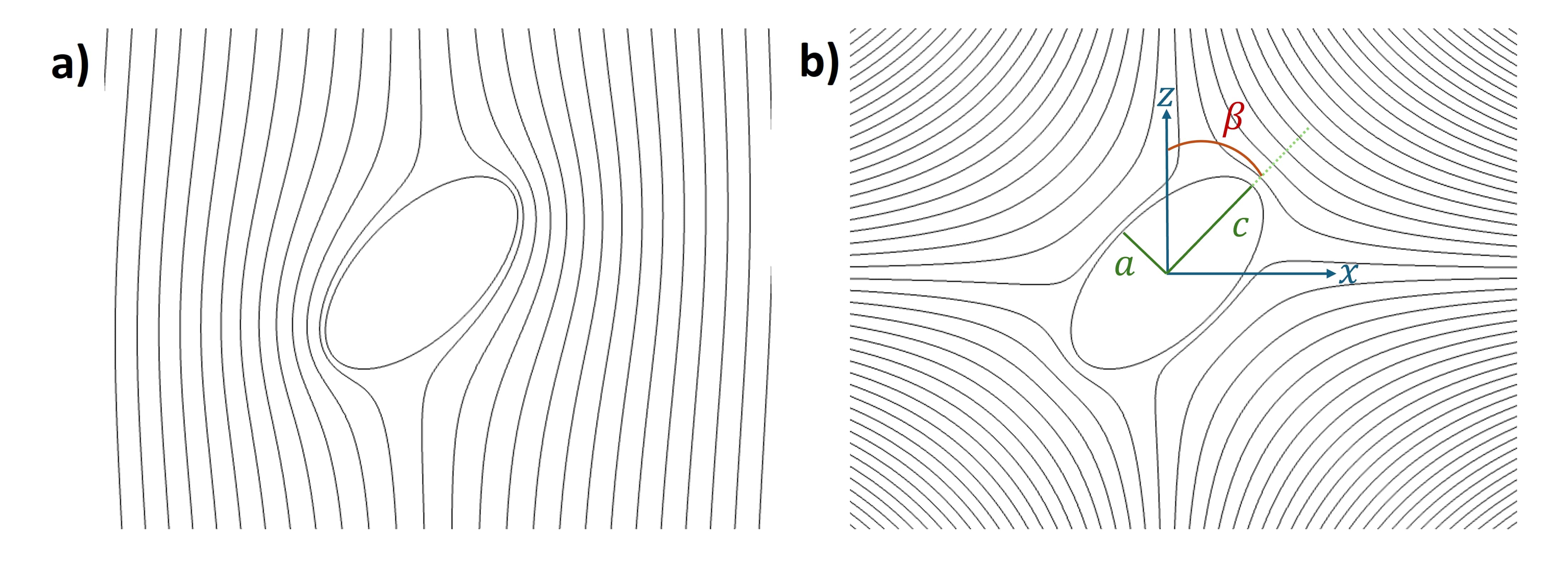}
        \caption{Magnetic field lines interacting with a diamagnetic ellipsoid, (a) for a uniform external magnetic field and (b) for an external quadrupolar magnetic field.}
        \label{fig:fieldlines}
\end{figure}

We consider $a,b$, and $c$ small enough with respect to $R$ so that in all situations, all the points of the ellipsoid will be on the region where Eq.~\eqref{eq:AHC} holds. 

To simplify the treatment we shall consider, unless explicitly indicated, that we have a perfectly diamagnetic ($\chi=-1$) ellipsoid with $a=b$ (a spheroid),  near the center of the trap. 

\subsection{Forces after translations}
\label{subsec:forces}

The stable position for translation of a levitated spheroid is when its center coincides with the center of the AHC, where the applied field is zero (we neglect gravity). Any translation of the ellipsoid from this position results in a restoring force that tends to push the ellipsoid to the center again. 
For $\chi=-1$ materials (ideal superconductor), some currents are induced in its surface to have ${\bf B}=0$ inside it (see Fig.\ref{fig:fieldlines}).

In this case, the magnetization inside the ellipsoid is not uniform. Since we are considering a region where the applied field is linear, if the center of the ellipsoid is moved to a point ${\bf r}_0$, we have
\begin{equation}
    \Ha ({\bf r}) = \Ha({\bf r}_0) + \Ha({\bf r}'),
    \label{eq.lin}
\end{equation}
where ${\bf r}'$ points to an ellipsoid's point \textit{as if it were located at the origin}.

As we have assumed linear materials, the magnetization of the spheroid will be the sum of the magnetization induced by these two fields. The currents induced by the second term will have a complicated distribution but, because of the symmetry, the forces they receive sum up to zero when integrated all over the spheroid's surface. 

Thus, \textit{as for the force evaluation is concerned}, we can consider only the currents induced because of the field given by the first term of the right-hand side of Eq.~\eqref{eq.lin}. That is, the force exerted over the ellipsoid with its center at ${\bf r}_0$ can then be written, using \eqref{eq:MVsHa}, as 
\begin{equation}
    {\bf F}=\mu_0 V ({\bf M}\cdot\nabla)\Ha\, =\mu_0 V (\SS \Ha\cdot\nabla)\Ha,
    \label{eq:force.dipole}
\end{equation}
where applied field and its derivatives should be evaluated at ${\bf r}_0$, the "displaced" center of the spheroid. 

The trapping frequencies are then evaluated from the variations of the force components with respect to the coordinates ($x_l=x, y$, or $z$):
\begin{equation}
    \omega_{kl} = \sqrt{\frac{1}{m} \left(-\left.\frac{\partial F_k}{\partial x_l} \right|_{x_l\rightarrow 0} \right)},
\end{equation}
being $m$ the mass of the ellipsoid. Using Eqs.\eqref{eq:AHC} and \eqref{eq:force.dipole} one gets
\begin{eqnarray}
   \omega_{xx} &=& \omega_0 \sqrt{-\SS_{xx}}, \label{eq:omega11} \\ 
   \omega_{yy} &=& \omega_0 \sqrt{-\SS_{yy}}, \label{eq:omega22} \\ 
   \omega_{zz} &=& 2 \omega_0 \sqrt{-\SS_{zz}}, \label{eq:omega33} 
\end{eqnarray}
where $\omega_0$ is defined as 
\begin{equation}
    \omega_0=\frac{H_0}{R} \sqrt{\frac{\mu_0}{\varrho}},
    \label{eq:omega0}
\end{equation}
being $\varrho$ the mass density of the ellipsoid. When $\beta=0$ or $\beta=\pi/2$, $\omega_{kl} = 0$ for all $k\neq l$. For an ideal superconducting sphere, Eqs. \eqref{eq:omega11}, \eqref{eq:omega22} and \eqref{eq:omega33} reduce to $\omega_{xx}=\omega_{yy}=(1/2)\omega_{zz}=\omega_0\sqrt{3/2}$, which coincides with Ref. \cite{Romero-Isart2012}. Considering a lead levitating particle, $\varrho=1.09\cdot 10^{4}$ kg/m$^3$, in an AHC coil of $\mu_0H_0/R\simeq$ 75 T/m \cite{Hofer2023}, we obtain $\omega_0\simeq 640$ rad s$^{-1} \simeq 2\pi \times 100$ Hz.

\subsection{Comparison with cylinders and cuboids}
\label{sec.comparison}

Up to now, we have considered the levitated superconductor with an ellipsoidal shape. It is clear that this is an excellent case for extracting analytical expressions and it represents a pretty good approximation for other shapes. However, the fabrication of such samples is more complicated than other "simpler" shapes such as cylinders or cuboids. In this section we present numerical results of trapping frequencies, comparing them with the analytically obtained for ellipsoids.

To compare the different geometries, we consider spheroids with $a=b$, and compare them with cylinders of radius $a$ and length $2c$ and cuboids with a square cross-section of sides $2a$ and length $2c$. Note that the volume of the three samples is not the same: $V_{\rm ell}=4\pi a^2 c /3$, $V_{\rm cyl}=2 \pi a^2 c$, and $V_{\rm cub}=8 a^2 c$ for the ellipsoid, cylinder, and cuboid, respectively. The mass density $\varrho$ is considered the same. To compare different aspect ratios, we have fixed the value of $a$ and varied the value of $c$. In all the cases, we have considered that all the samples are in the region where the approximation of the quadrupolar external field holds. 

\begin{figure}
	\centering
        \includegraphics[width=0.48\textwidth]{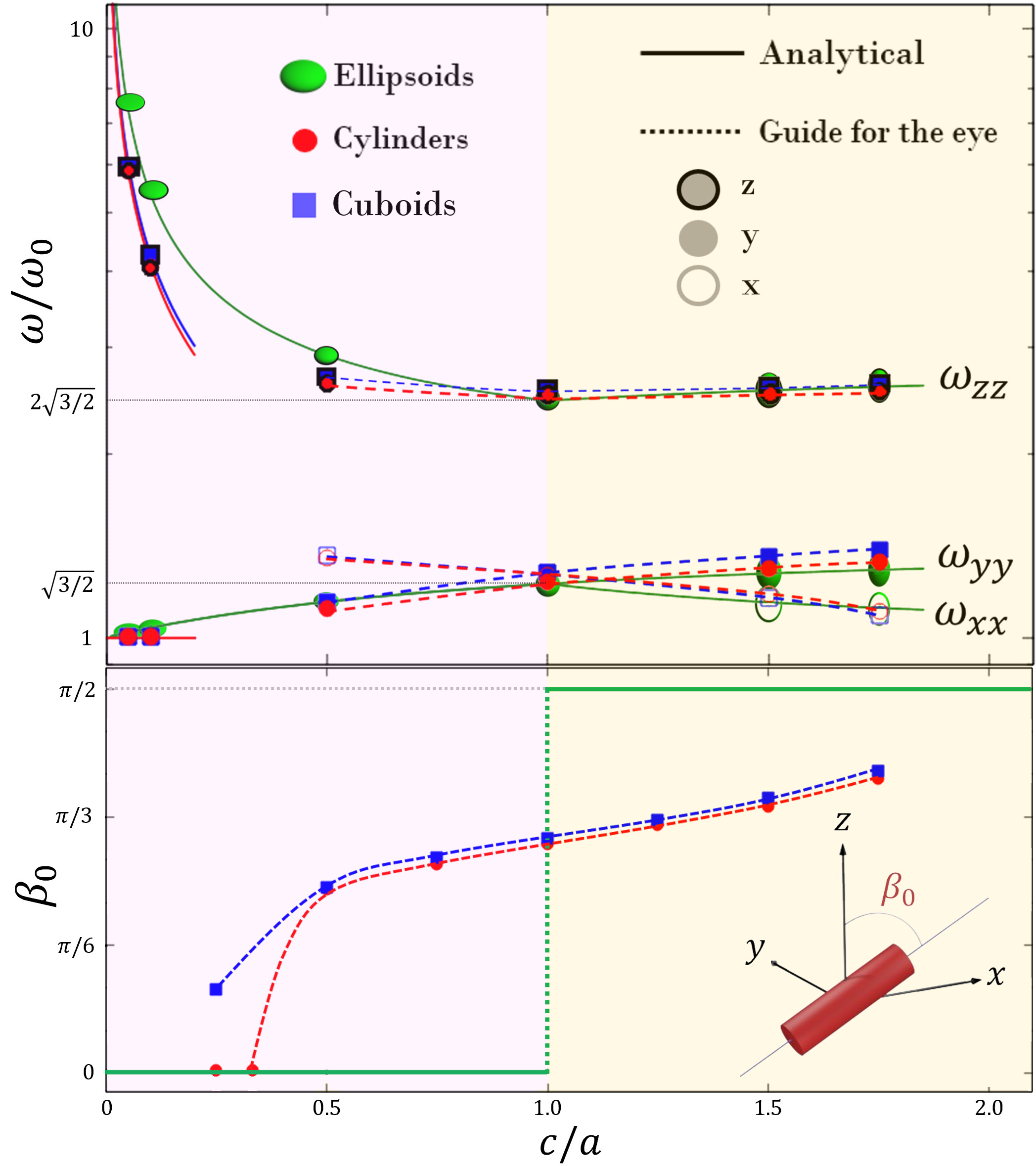}
	\caption{On top, there are the normalized translational frequencies' components (in logarithmic scale) as a function of $c/a$ for ellipsoids (green ellipsoids), cylinders (red circles), and cuboids (blue squares). Each point is evaluated from the corresponding stable-rotation angle $\beta_0$. For the same aspect ratio, this stabilization angle is different depending on the geometry. On the bottom, we plot the numerically evaluated stabilization angle as a function of the aspect ratio. Therefore, all frequencies are evaluated with $\beta=\beta_0$ (the corresponding rotational-stability angle) and $\chi=-1$. The lines for the ellipsoids correspond to the analytical expressions [see Eqs.  \eqref{eq:omega11}, \eqref{eq:omega22}, and \eqref{eq:omega33}]. For the cylinders and cuboids, the dotted lines are guides for the eye, and the solid lines are the analytical expression for the small $c/a$ limits [Eqs.\eqref{eq:limitthin} and \eqref{eq:limitthinx}]. The purple region corresponds to oblate ellipsoids and the orange one to prolate ellipsoids. All the dots are numerically calculated values using finite-element methods. The inset below is a scheme of the definition of $\beta_0$ using a cylinder as an example.} \label{fig:comparison}
\end{figure}

In Fig. \ref{fig:comparison} we show the calculated frequency for the three shapes as a function of the aspect ratio. It is important to note that, as we shall see in section \ref{subsec.torques}, the equilibrium angle $\beta_0$ with respect to rotation varies when $c/a$ changes. In Fig. \ref{fig:comparison} we show the calculated translational frequencies considering $\beta=\beta_0$ (shown in Fig. \ref{fig:comparison} for cylinders, cuboids, and ellipsoids). We have also calculated some values for the spheroid to double-check the previous equations and to ensure that the numerical imprecision of the calculations does not affect our results. The values for cylinders have also been double-checked in Ref. \cite{private}.

The main facts we observe are: (i) the equilibrium angle for stability, $\beta_0$, is different depending on the sample and its aspect ratio $c/a$; (ii) the evaluated vertical frequencies are larger than the horizontal ones for all the considered geometries; (iii) there is a sudden change in $\beta_0$ for the ellipsoids when $c/a=1$ yielding a kink in the ellipsoid's plot; (iv) for oblate ellipsoids ($c/a$<1), $\omega_{xx}=\omega_{yy}$ because $\beta_0=0$; for prolate ellipsoids ($c/a>1$), $\omega_{xx}\neq\omega_{yy}$ because $\beta_0=\pi/2$; (v) the vertical and horizontal frequencies for cylinders and cuboids are very similar in this comparison; (vi)  the frequencies become greater as more fraction of the superconducting volume is located closer to the $z=0$ plane \textit{and/or} the $z$-axis (suggesting a $z-$revolution astroid-like shape \cite{Lawrence1972book} for the levitated particle).

It is interesting to study the $c\ll a$ limit (very thin samples). It is known that the external susceptibility at small applied fields is $\chi_0=f_0 a/c$ with $f_0=4/(3\pi)$ for cylinders \cite{Clem1994} and $f_0=0.9094$ (numerically found value) for cuboids \cite{Chen2008}. The resulting vertical and horizontal frequencies are 
\begin{eqnarray}
 \omega_{zz} &=& 2\omega_0\sqrt{f_0\frac{a}{c}}\, , \label{eq:limitthin}\\
    \omega_{xx} &=& \omega_{yy} = \omega_0 \label{eq:limitthinx}\, .
\end{eqnarray}
Note that, in this limit, $\beta_0=0$ and $\omega_{xx}$ and $\omega_{yy}$ are the same for both samples. For thin oblate spheroids, one can also find that $\omega_{zz} \propto \sqrt{a/c}$ and $\omega_{xx}$ ($=\omega_{yy}$) tends to $\omega_0$.


\subsection{Torques after rotations}
\label{subsec.torques}

Even if the center-of-mass of the ellipsoid is not displaced from the origin of coordinates, the spheroid can rotate. Rotation of microscale diamagnets can exhibit interesting quantum phenomena \cite{Stickler2021}. In general, the torque on a given magnetized body due to an external field is given by
\begin{equation}
    {\bf T} = \mu_0 \int_V  {\bf r} \times (\M \cdot \nabla) \Ha \,\, {\rm d}V.
    \label{eq:tdef}
\end{equation}

When the spheroid is not rotated (their principal axes point along the coordinates axes) the total torque is zero, although the sample is magnetized. To simplify the treatment of librations we consider that the rotation is over the $y$-axis, represented by the angle $\beta$. We can define the librational frequency of the trap as
\begin{equation}
\omega_\beta= \sqrt{\frac{1}{I_y} \left(- \left.\frac{\partial T_y}{\partial \beta}\right|_{\beta\rightarrow \beta_0}\right)},
\label{eq:wangular}
\end{equation}
where $I_y$ is the moment of inertia of the spheroid with respect to the $y$-axis, $I=(1/5) m (a^2+c^2) = (4/15) \pi a^2 c (a^2+c^2) \varrho $ for a solid spheroid.

$\beta_0$ is the angle with respect to the $z$-axis at which the total torque cancels when only libration is considered. Thus, it represents the rotational equilibrium angle. Ellipsoids, cylinders, and cuboids with identical $c/a$ ratios will stabilize libration at different $\beta_0$ angles (see Fig. \ref{fig:comparison}). For cuboids, there would also be an $\alpha$ or $\gamma$ dependency, that has not been considered here.

In Ref. \cite{Karma2014}, the angular frequency was expressed as $\omega_\beta= \frac{c}{a\sqrt{(a/c)^2+1}} \sqrt{\frac{ 15}{4\pi \varrho} k_\beta}$, where $k_\beta=-\frac{1}{c^2}\left.\frac{\partial T_y}{\partial \beta} \right|_{\beta\rightarrow 0}$. $k_\beta$ was been defined as the effective spring constant of an effective librational oscillator of length $c$. This definition is similar to ours, although adapted to the vibration of a cantilever.

\begin{figure}
    \centering
    \includegraphics[width=0.48\textwidth]{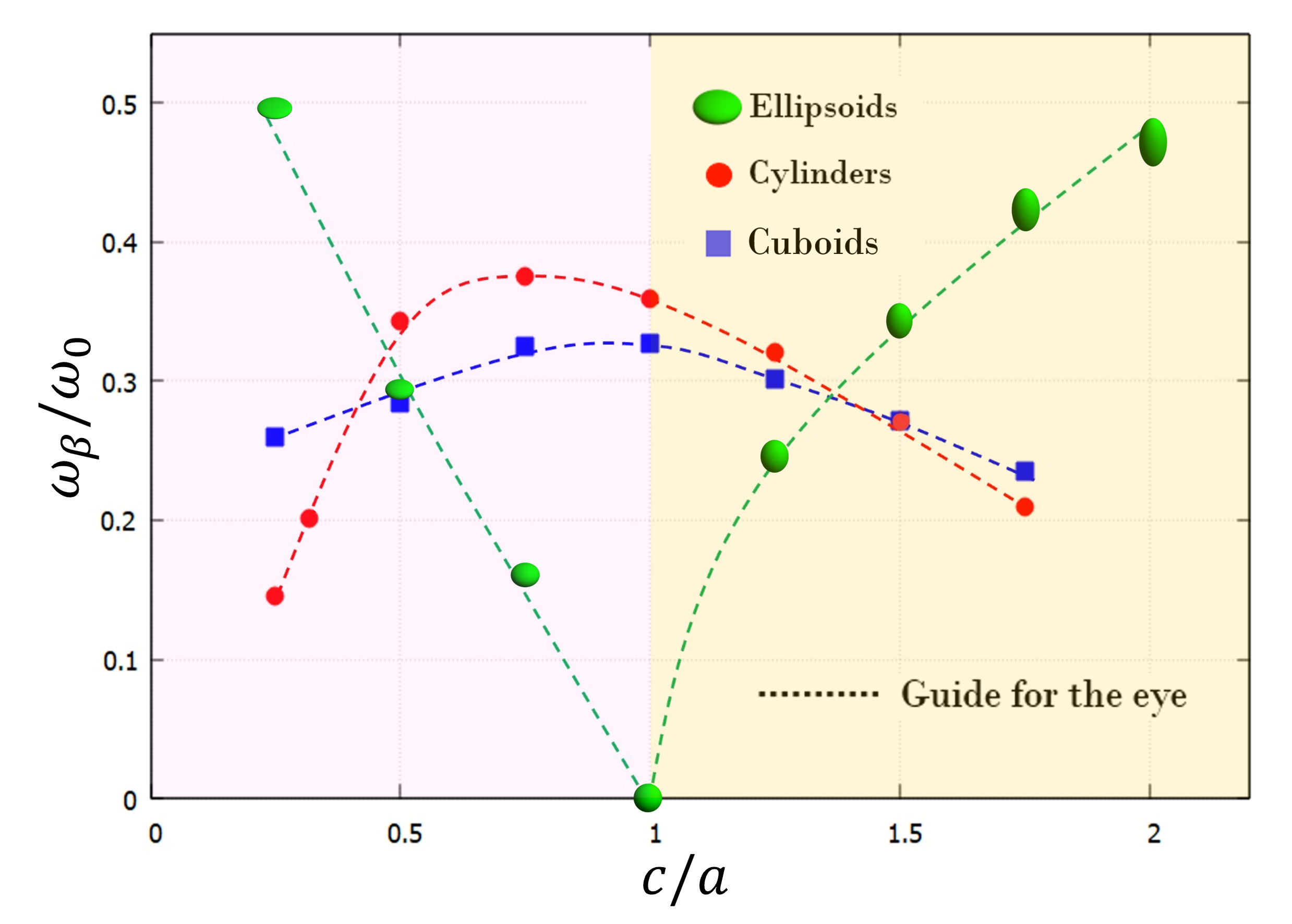}
    \caption{Librational frequencies for angular vibration of ellipsoids (green), cylinders (red), and cuboids (blue), as a function of the aspect ratio $c/a$. Each point is evaluated from the corresponding stable-rotation angle $\beta_0$, see Fig. \ref{fig:comparison}. The purple region corresponds to oblate ellipsoids and the orange one to prolate ellipsoids. Dashed lines are guides for the eye.}
    \label{fig:w_tor}
\end{figure}

Although the external field is known, after rotation, we cannot separate the external field in a non-torque-producing term plus a non-zero uniform term [as we did in Eq.~\eqref{eq.lin}]. In the case of a superconducting spheroid with $\chi=-1$, the reaction of the superconductor is the induction of surface currents ${\bf K}$ whose value is evaluated from the discontinuity of the tangential component of the $\B$-field at the (rotated) surface of the ellipsoid. In this case, the total torque can be also expressed as
\begin{equation}
    {\bf T} = \mu_0 \int_S {\bf r}\times ({\bf K} \times \Ha) \,{\rm d}S.
    \label{eq:torque.K}
\end{equation}
We could not find in this case an easy analytical expression for the currents. Approximately, they could be evaluated from the currents at the surface of a sphere \cite{Hofer2019} adequately distorted to account for the spheroidal shape and then evaluate the above integral. In any case, the expressions would not be simple enough as they are in terms of elliptical integrals. We present in Fig. \ref{fig:w_tor} the librational frequencies evaluated with Eq.~\eqref{eq:wangular}, after numerically evaluating the torques. The kink in the librational frequency for the ellipsoids is explained by the change in $\beta_0$ from 0 to $\pi/2$ when the ellipsoid passes from oblate to prolate, as shown in Fig. \ref{fig:comparison}.

We observe in Fig. \ref{fig:w_tor} that when $c/a$ is close to unity, cylinders, and cuboids are more rigid than spheroids, concerning libration (as expected, since for $c/a=1$ the ellipsoid is a sphere which can rotate freely, because of the symmetry). However, as the ellipsoid becomes more prolate or more oblate, it becomes more rigidly levitated. 

Finally, note that an already rotated ellipsoid, if moved laterally or vertically, would experience, apart from the evaluated restoring force and torque, extra forces that go in the direction perpendicular to the displacement, and, as a consequence, also torques in other directions [see Fig. \ref{fig:generalS} and Eq. \eqref{eq:SSspheroid}]. The general movement, taking into account all the possibilities is beyond the scope of this work, but we would like to note that many different possible modes of vibration can appear and that the particular geometry of the levitating sample can play a significant role in their trapping frequencies.

\section{Conclusions}
\label{sec.conclusions}
We have found the external susceptibility matrix, $\SS$, that directly relates the magnetization of the ellipsoid with the \textit{uniform applied} field, for all possible relative orientations between them. The key result is that we can express the applied field in the convenient coordinates needed for describing the magnetic system and the magnetization is found directly as a function of the demagnetizing factors, which are tabulated assuming that the field is along one of the principal axes of the ellipsoid. 

We have used this result to evaluate the forces received considering a $\chi=-1$ spheroid located at the central region of an anti-Helmholtz coil system. From these forces, the translational (analytically and numerically) and the librational (numerically) trapping frequencies have been evaluated. 

The results found, although derived in an idealized case, could be useful in the field of magnetomechanics, since the knowledge of analytical (despite approximate) equations can guide the design of the experimental systems. Moreover, the detailed description of the levitation system and the geometry effects can also serve to calibrate a given experiment and, thus, increase the performance of the experimental setup. 

In a more general scope, the described external susceptibility and its consequences could help in performing demagnetization corrections in a broad type of experiments since we have found the external susceptibility tensor as the relation between the internal magnetization (a measure of the reaction of the material) and the \textit{external} applied field (an easily controllable magnitude).

\begin{acknowledgments} 
We acknowledge financial support from (a) Spanish Ministry of Science and Innovation MCIN/AEI through PID2019-104670GB-I00; (b) Horizon Europe 2021-2027 Framework Programme (European Union) through the SUPERMEQ project (Grant Agreement number 101080143). J.C.-S. acknowledges funding from AGAUR-FI Joan Or\'o grants (2023 FI-3 00065), Generalitat de Catalunya. W.W. acknowledges funding in part by the Knut and Alice Wallenberg Foundation through a Wallenberg Academy Fellowship, the QuantERA project CMON-QSENS!, and the European Research Council under Grant No.~101087847 (ERC CoG SuperQLev). G.H.\ acknowledges support from the Swedish Research Council (Grant No.\ 2020-00381). This research was funded in whole or in part by the Austrian Science Fund (FWF) [10.55776/esp525]. 
\end{acknowledgments}


%

\end{document}